\documentclass[review,number,sort&compress]{elsarticle}

\usepackage{lineno}
\usepackage{color}

\usepackage[section]{placeins}

\journal{Nuclear Instruments and Methods}

\begin{document}

\begin{frontmatter}
\date{\today}

\title{Design, Modeling and Testing of the Askaryan Radio Array South Pole Autonomous Renewable Power Stations}

\author[kupa,mephi]{D.Z.~Besson}
\ead{zedlam@ku.edu}
\author[kupa]{D.M.~Kennedy\corref{cor1}}
\ead{dmkennedy@ku.edu}
\author[idl]{K.~Ratzlaff}
\ead{ratzlaff@ku.edu}
\author[idl]{R.~Young}
\ead{rwyoung@ku.edu}

\cortext[cor1]{Corresponding Author: Daniel Kennedy, Department of Physics and Astronomy, University of Kansas, 1082 Malott Hall, 1251 Wescoe Hall Drive, Lawrence, KS 66045-7582; Email, dmkennedy@ku.edu; Phone, 785-864-4626.}

\address[kupa]{Department of Physics and Astronomy, University of Kansas, 1082 Malott Hall, 1251 Wescoe Hall Drive, Lawrence, KS 66045-7582}
\address[mephi]{Moscow Engineering and Physics Institute, 31 Kashirskaya Highway, Moscow 115409, Russia}
\address[idl]{Instrumentation Design Laboratory, University of Kansas, 6042 Malott Hall, 1251 Wescoe Hall Drive, Lawrence, Kansas 66045-7582}

\begin{abstract}
We describe the design, construction and operation of the Askaryan Radio Array (ARA) Autonomous Renewable Power Stations, initially installed at the South Pole in December, 2010 with the goal of providing an independently operating 100 W power source capable of year-round operation in extreme environments. In addition to particle astrophysics applications at the South Pole, such a station can easily be, and has since been, extended to operation elsewhere, as described herein.
\end{abstract}


\end{frontmatter}

\pagebreak

\section{Introduction}
\begin{figure}[h!]
\centering
\includegraphics[width=0.75\linewidth]{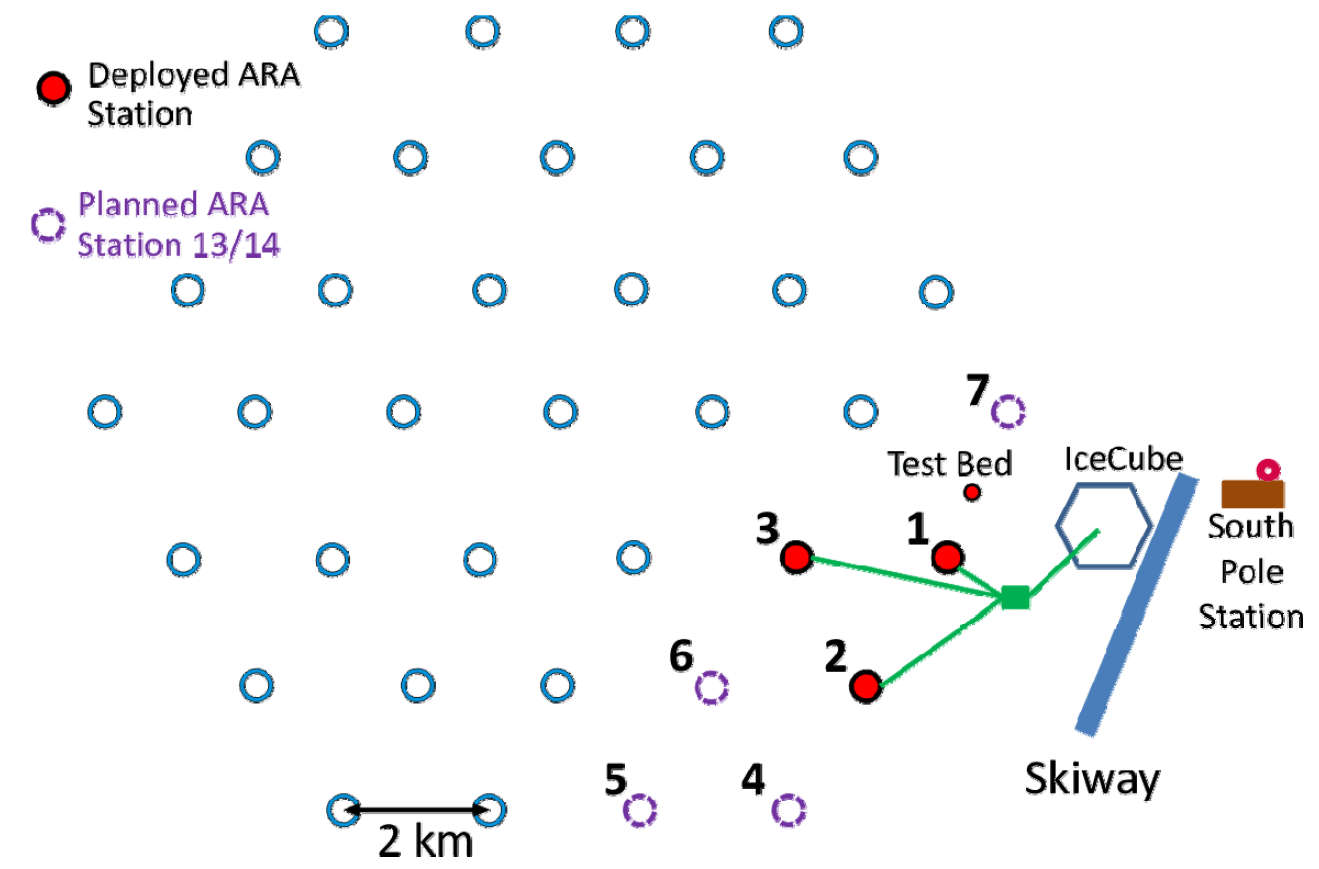}
\caption{Proposed ARA station map. Phase one of the radio array is designed to cover a $200km^2$ area adjacent to South Pole Station. The long distances between station nodes was a driving factor in the power station design. \label{fig:ara_map}}
\end{figure}

Research was initiated in 2009 for the development of a remote power station to support the Askaryan Radio Array South (ARA) Pole neutrino detector\cite{Allison}. The purpose of the array is to detect ultra high energy cosmic neutrinos using the Askaryan effect\cite{Askaryan61}. When neutrinos with energies above $10^{18} eV$ interact with the ice, a cascading particle shower occurs, resulting in coherent Cherenkov radio emission at wavelengths greater than the transverse scale of the developing cascade. For showers developing in ice, this scale is set by the Moliere radius and corresponds to approximately 20 cm. Phase one of the radio array is designed to cover a $200km^2$ area adjacent to South Pole Station (Fig. \ref{fig:ara_map}), and, as of this writing, is approximately 10\% complete. The polar ice has extremely high radio transparency, allowing the radio signals to be detected over the large size of the array. To maximize sensitivity, $2 km$ spacing was selected \cite{Allison}. The long distances between station nodes was a driving factor in the power station design.

\section{Design Requirements}
\subsection{Detector Power}
Each station (a.k.a. ``detector node'') requires approximately 120W of continuous power with a target of at least 95\% up-time to maximize neutrino sensitivity. For detector nodes located near South Pole Station, entrenched power lines can be connected directly to the station. For remote nodes, power can be provided by direct line, diesel or jet-fuel generator, fuel cell, wind turbine, and solar power.

\subsection{Environmental Effects}
For direct power connections to South Pole Station, two inch wide, one foot deep trenches must be created for the cables, then covered, to prevent possible collisions between vehicles and the cables. Due to the limited logistical support and significant $I^2R$ losses, this option is increasingly
challenging for distant nodes.

Autonomous diesel or jet-fuel generators could be constructed at each node location. This method is currently in use at the University of New South Wales' Automated Astronomical Site-Testing International Observatory (AASTINO) at Dome C. This location uses a computer-controlled jet-fuel Stirling engine \cite{domec}. Although this option offers simplicity and is fairly reliable, it also requires transport of large quantities of fuel each year and was therefore not considered for prototyping at South Pole.

Another option is renewable power production using a combination of wind turbines and photo-voltaic (PV) panels. This option is highly dependent on the environment conditions at deployment site. Wind speeds at the South Pole average between 4 and 8 m/s depending on height above the surface; these wind speed values are just above the typical threshold, or ``cut-in'' speeds of current wind turbines. Although solar power is only available for half of the year, the cold temperatures result in higher panel efficiency during the operating period \cite{efficiency}. Consequently, the use of an intermittent power supply (either solar or wind) would require redundant power sources as well as battery buffering at each node to reach the required station up-time. The combination of PV and wind turbines was selected for initial deployment, given the ultimate goal of a $10^3 km^2$ aereal scale radio array.

\subsection{Wind Turbine Selection}
\begin{figure}[h!]
\begin{picture}(350, 175)(0,0)
\put(0,0){\includegraphics[scale=0.75]{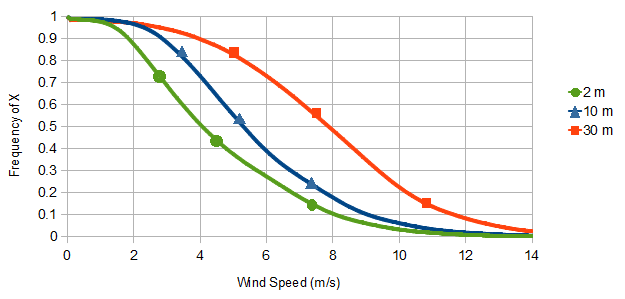}}
\end{picture}
\caption{NOAA integral wind speed frequency distribution. Shown is the fraction of times a given wind speed value is exceeded.}
\label{fig:freq_dist}
\end{figure}

Prior to selecting wind turbines, an understanding of the wind profile at the South Pole was required. Near surface wind observation data was collected from measurements by the National Oceanic and Atmospheric Administration (NOAA) / Oceanic and Atmospheric Research Earth System Research laboratory (OARES) / Global Monitoring Division (GMD) observatory at Amundsen-Scott South Pole Station. This included per minute wind speed and direction measurements at heights above the surface of 2m, 10m, and 30m over several years. Figure (\ref{fig:freq_dist}) illustrates the frequency distribution of wind speeds at the different anemometer heights. This Figure shows the fraction of times that the wind speed exceeds a given value, and can therefore be translated into a duty cycle for a given turbine, knowing the cut-in speed for that model of turbine.

Research into wind turbines was started in 2010 with selection based on mass, start-up speed, price, rated power, and rated cut-in wind speed. Table (\ref{tb:turbs}) lists the selected candidate turbine specifications.

\begin{table}[h]
\centering
\caption{Specifications for wind turbines considered for prototyping at South Pole.} \label{tb:turbs}
\begin{tabular}{| c | c | c | c | c | c |}
\hline
Model & Power & Mass & Startup & Rated & Season \\
 & (Watts) & (kg) & Speed (m/s) & Speed (m/s) & \\
\hline
\hline
Bergey XL1 & 1000 & 34 & 3 & 11 & 2010-2011 \\
Hummer & 1000 & 15 & 3 & 9 & 2010-2011 \\
Raum Energy & 1500 & 39 & 3 & 11 & 2010-2011 \\
\hline
Whisper & 900 & 21 & 3 & 12.5 & 2011-2012 \\
Aero6gen-F & 240 & 16 & 3 & 21 & 2011-2012 \\
\hline
\end{tabular}
\end{table}

\subsection{Tower Selection}
Multiple types of wind turbine towers were selected for initial deployment. These were selected based on height, erection method and logistical support requirements. All towers utilized a tilt-up system with guyed wires.

\begin{figure}[h!]
\centering
\includegraphics[width=0.6\linewidth]{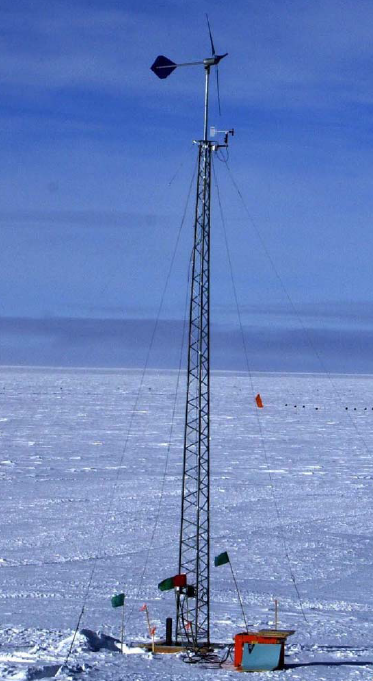}
\caption{50' lattice tower used for Raum turbine. Minimal logistical support at the South Pole required that the lattice tower be constructed prior to shipping and sent in sections. \label{fig:lattice}}
\end{figure}

A 50 foot lattice tower from Raum Energy was used for the Raum 1.5 kw turbine and utilized a single set of guy wires (Fig. \ref{fig:lattice}). Minimal logistical support at the South Pole required that the lattice tower be constructed prior to shipping and sent in sections. This proved very inconvenient for shipping. Moreover, the amount of structural support provided by the lattice design was not necessary for the site conditions.

\begin{figure}[h!]
\centering
\includegraphics[width=0.75\linewidth]{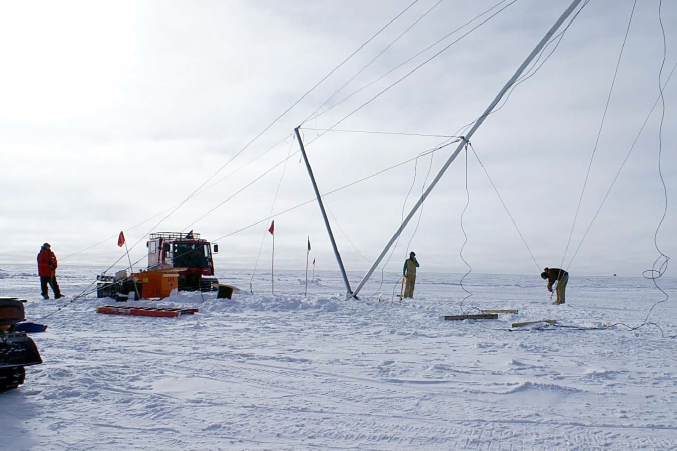}
\caption{60' tilt tower with kevlar guys. The Kevlar guys were used in place of steel to reduce static build-up, however no static build-up was noted on any of the towers after deployment. \label{fig:tilt}}
\end{figure}

The Bergey XL1 turbine used a 60 foot monopole provided by Bergey Windpower. This tower utilized three sets of three kevlar guy wires. Once the wires were adjusted, the tower was very stable, however this required significant time and effort in the cold. The Kevlar guys were used in place of steel to reduce static build-up, however no static build-up was noted on any of the towers after deployment (Fig. \ref{fig:tilt}). A 60 foot monopole with three guy wires was used for the Hummer turbine. This tower was deemed too unstable and subsequently reduced to 50 feet.

\begin{figure}[h!]
\centering
\includegraphics[width=0.75\linewidth]{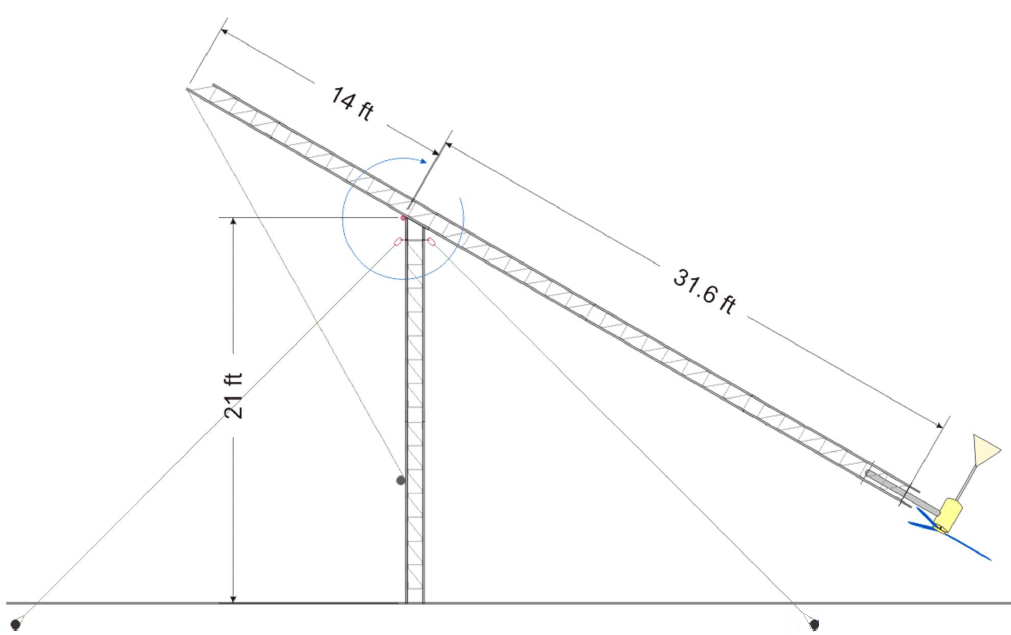}
\caption{Diagram of a Rohn fold-over tower planned for construction on future deployments enabling more efficient construction and turbine maintenance. \label{fig:fold}}
\end{figure}

A custom tilt 20 foot tower with gin pole was developed at the University of Kansas and deployed during the second season, in conjunction with the Aero6gen turbine, without issue. A Rohn fold-over tower is planned for construction on future deployments enabling more efficient construction and turbine maintenance (Fig. \ref{fig:fold}).

\subsection{Tower Height and Surface Roughness}
\begin{figure}[h!]
\centering
\includegraphics[width=0.9\linewidth]{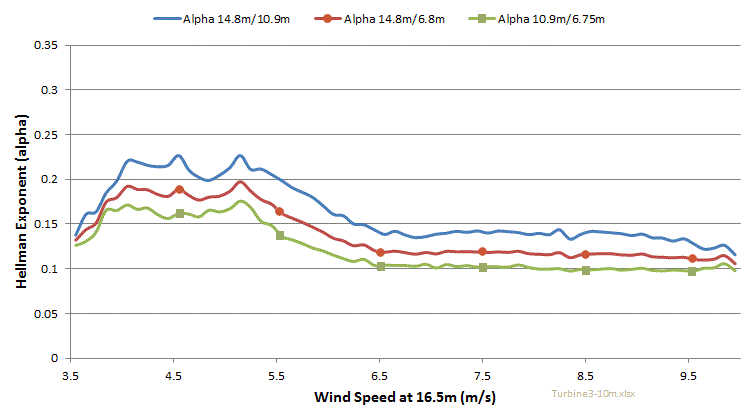}
\caption{The Hellman Exponent ($\alpha$), shown as a function of height above the surface, is a measure of the significance of height on the wind speed. In general, larger heights favor higher wind speeds as the distance relative to turbulent near-surface layers increases. \label{fig:hellman}}
\end{figure}

The required tower height is directly related to the surface roughness on the plateau.  The Hellman Exponent, or $\alpha$, is a measure of the significance of height on the wind speed. In general, larger heights favor higher wind speeds as the distance relative to turbulent near-surface layers increases. The parameter $\alpha$ is defined as:

\begin{equation}
\alpha = ln(V/V_0) / ln(H/H_0)
\end{equation}

where $V$ and $V_0$ represent wind speeds measured at heights $H$ and $H_0$ \cite{Kaltschmitt}. If the surface were glassy smooth, airflow would be near-laminar, resulting in small values of $\alpha$ ($\sim$0.06). In proximity to surface structures, $\alpha$ can rise to 0.5 or greater. As $\alpha$ increases, it becomes increasingly important to build the tower as high as practically achievable. The measurement of $\alpha$ for multiple heights on the Antarctic plateau is shown in Figure (\ref{fig:hellman}).

This measurement clearly demonstrates that the wind on the plateau is affected by surface roughness, due to natural ice structures known as  ``sastrugi''.

\subsection{Photo-voltaic Panel Selection}
Wind power is augmented by solar power for approximately 6 months between mid-September and mid-March. Outside of this period, the sun is below the Antarctic horizon. Research into candidate PV panels was conducted in 2010 resulting in the selection of the Sharp NU-U235F1 325 Watt panel. The first PV panel was installed during the second year of ARA deployment, the 2011-2012 season.

\begin{figure}[h!]
\centering
\includegraphics[width=0.8\linewidth]{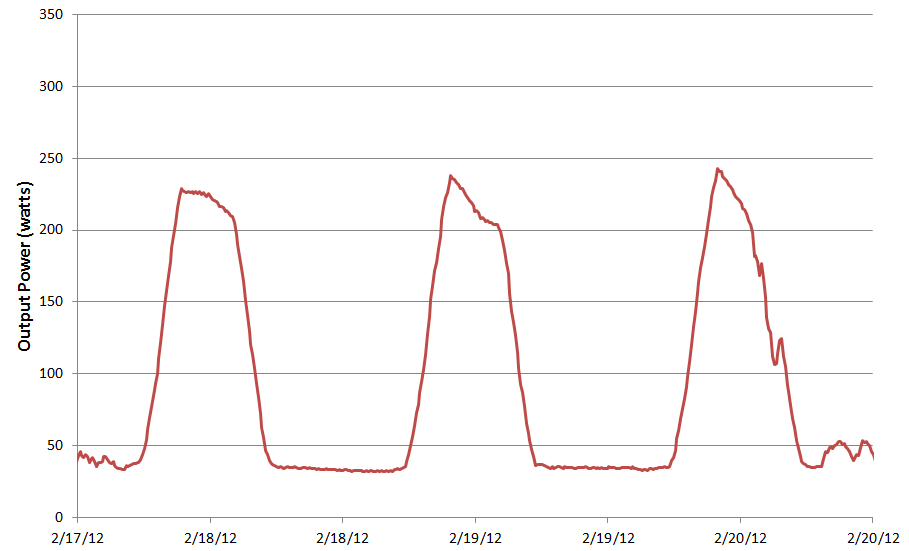}
\caption{PV panel output (325W panel) shown over three days. The vertical mounting takes advantage of the albedo of the snow resulting in continuous power output even while the sun is behind the panel. \label{fig:pv_cycle}}
\end{figure}

\begin{figure}[h!]
\centering
\includegraphics[width=0.8\linewidth]{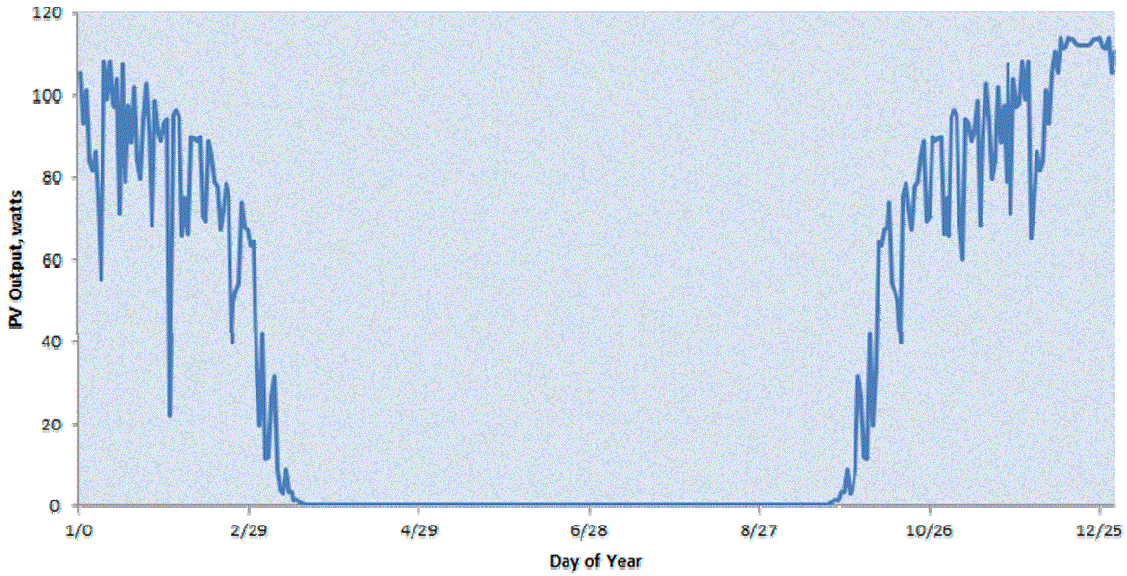}
\caption{Average PV output (325W panel). Wind power is augmented by solar power for approximately 6 months between mid-September and mid-March. Outside of this period, the sun is below the Antarctic horizon. \label{fig:average_pv}}
\end{figure}

A vertical PV tower was designed to take advantage of the albedo of the snow. The panel was oriented vertically, such that the angle was equal to the latitude, 90 degrees, resulting in highest efficiency when the sun is at the horizon. This allows the panel to operate without sun tracking. Figure (\ref{fig:pv_cycle}) shows the power output over three days. The power output remains above zero even while the sun is behind the panel. The measured average power output from the initial PV panel is shown in Figure (\ref{fig:average_pv}).

The tower was designed with sufficient ground clearance to allow snow buildup between seasons without impeding panel operation. Measurements have shown that the tower base has received in excess of one foot per year of snow buildup. Deployment of a second panel on the reverse side of the tower is planned.

\subsection{Battery Selection}
Two 12 volt DEKA 8G31DT gel batteries were used for energy storage at each turbine. This battery model was previously characterized by UNAVCO \cite{unavco}. These batteries are particularly suited to deep charge/discharge cycles, emergency backup systems, and unusually demanding systems such as marine and off-road vehicles \cite{deka}.

\subsection{Sensor Selection}
TMP125 model temperature sensors were placed inside and immediately outside the instrument housing box as well as on the tower. These sensors measured SHM box temperature, outside temperature and free-air temperature respectively. Although these sensors are rated to -40C, testing at temperatures as low as -70C resulted in no detectable errors.

RMYoung anemometers (propeller-type) were placed near the top of each tower to independently measure wind speed. On one of the 60 foot towers, two additional sensors were placed at intermediate heights. This type of anemometer is routinely used on the polar plateau by UNAVCO. They consume little power, especially compared with solid-state units, and have proven to be durable in the harsh Antarctic climate. RHYoung anemometers are also used for the Automatic Weather Stations Project by the Antarctic Meteorological Research Center at UW-Madison.

The environmental measurements have several goals. Temperature readings are used to assess box heating requirements. Wind speed is used to compare with turbine power for evaluating generator performance. Additionally, readings from multiple elevations are used for determining wind speed as a function of tower height. This also measures the effect of the sastrugi on turbulence, which, in turn, affects wind speed.

\subsection{Instrument Housing}
\begin{figure}[h!]
\centering
\includegraphics[width=0.75\linewidth]{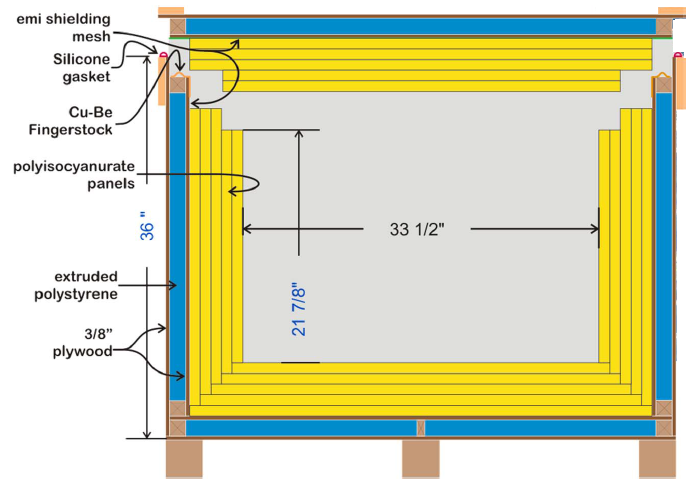}
\caption{Instrument box containing the system health monitoring board, batteries, and temperature sensors. The primary purpose of this box is to protect vital electronics from the environment. \label{fig:box}}
\end{figure}

\begin{figure}[h!]
\centering
\includegraphics[width=0.7\linewidth]{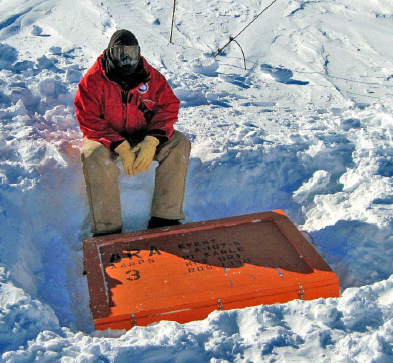}
\caption{Instrument box installed. The boxes were buried just below the ice surface, helping to shield the system from extreme temperatures. This resulted in added difficulty when performing maintenance or troubleshooting by requiring ice and snow removal for access. \label{fig:box_snow}}
\end{figure}

The instrument housing box (Fig. \ref{fig:box}) contains the system health monitoring board, 2 batteries, and temperature sensors. The primary purpose of this box is to protect vital electronics from the environment. The boxes were buried just below the ice surface, helping to shield the system from extreme temperatures and avoid snow drifting. This resulted in added difficulty when performing maintenance or troubleshooting by requiring ice and snow removal for access. However, a box on the surface would be equally difficult to access due to the development of large snow drifts (Fig. \ref{fig:box_snow}).

Plywood panels provide the structural strength while polyisocyanurate panels are used for insulation. EMI shielding mesh is placed between the plywood and insulation to shield radio frequency emission caused by the electronic switching circuitry. Measurements prior to deployment indicated that this mesh achieved approximately 40 dB RF isolation. Lastly, a silicone gasket is provided to seal the removable lid. Extra space was provided in the box to facilitate routing stiff cabling and allow battery bank expansion.

\subsection{Transmission Cables}
IMSA cables, which are designed for applications such as traffic lights, were used for power transmission lines.  The polyethylene insulation jacket provided reasonable flexibility at low temperatures. During the first season, thermoplastic rubber (seoprene) was used which became excessively stiff during station construction. This resulted in significant difficulty uncoiling cables and installing them in trenches. Initially, cabling was used from the IceCube Laboratory (see Figure 1) using RS485 for noise immunity. Ethernet was later used over fiber and CAT6.

\subsection{System Health Monitor}
\begin{figure}[h!]
\centering
\includegraphics[width=\linewidth]{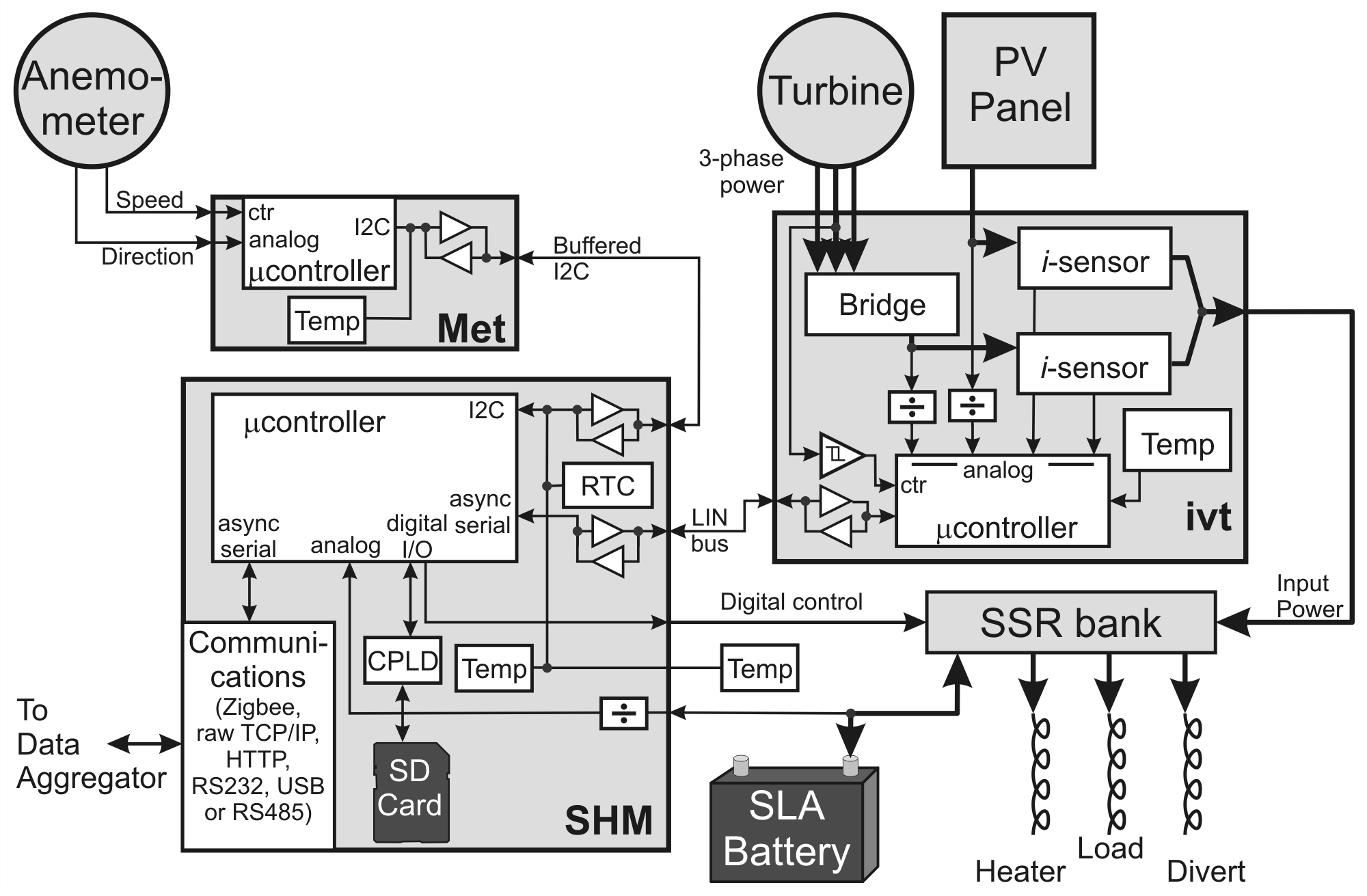}
\caption{Block diagram of the System Health Monitor (SHM) that controls and monitors the overall system. \label{fig:shm}}
\end{figure}

The System Health Monitor (SHM) was developed to control and monitor the overall system (Fig. \ref{fig:shm}). This system also replaces the function of a standard charge controller while eliminating electronic switching to reduce emissions. The monitoring functions are described in the following sections.

\subsection{Input Power}
Input power is measured for each turbine and solar panel with the Current-Voltage-Temperature (IVT) board. Depending on the turbine, power is rectified to DC either in the turbine nacelle or on the SHM. Rectification on the SHM is preferred because it allows the IV heat from the Schottky bridge diodes to be captured inside the power box; additionally, one phase may be used to directly measure the blade rotational frequency.

Input voltages are divided and buffered then digitized by the IVT microcontroller.  Hall-effect solid-state sensors (ACS714) detect the current and the resultant signal is also digitized by the microcontroller. The sensitivity of the current sensor has temperature dependency; consequently, a semiconductor temperature sensor (TMP175) is also mounted on the board. Although these temperature sensors are only rated to -40C, we have found them to be reliable as low as -70C.

The IVT microcontroller makes multiple measurements during the measurement period.
The IVT board communicates with the SHM via a LIN bus interface, allowing multiple IVT boards to be deployed in a single system.

\subsection{Wind and Temperature}
Wind and temperature measurements are sensed at up to 3 positions on the tower and used for tower height studies. These measurements are then summed and transmitted by the Metrology (Met) board located adjacent to the RM Young 05103 anemometers on the tower. This anemometer was selected for its capability for low-power operation and a good record of operation on the Antarctic plateau. The boards provide interfaces to the anemometers and to temperature sensors (TMP175). The Met board's microcontroller sums multiple measurements during the measurement period. A buffered I2C interface facilitates SHM communication with multiple Met boards.

\subsection{Power Distribution}
The power from the IVT board is distributed by a bank of Solid State Relays (SSRs).  Power can be directed to the battery, the load, a heater inside the instrument box, and a divert resistor outside the box. The switching logic comes from the SHM board.

\subsection{Communications}
The System Health Monitor board handles overall control and communications.

a. Data are received from the IVT and Met boards by standard protocols. The Met board(s) employ I2C, a standard synchronous serial protocol; boosters allow I2C to be used over longer than normal distances. The LIN bus, used to communicate with the IVT board(s), is an asynchronous serial protocol used in automotive applications; besides being robust, it can also be shared among multiple boards. The data are passed as comma-delimited ASCII strings, to facilitate debugging.

b. The battery voltage is divided down to the range of the microcontroller's analog input.

c. An SD card is interfaced via a CPLD which simplifies the interface. The SD card must be 2Gb or smaller.

d. Temperature is measured both on the board and off the board, for monitoring battery temperature and ice temperature.

e. An independent Real-Time Clock (RTC) keeps track of Time-of-Day; alkaline cell power backup carries it through possible power outages.

f. The Solid-State Relays (SSR) can be controlled directly from the microcontroller digital outputs since the SSRs have integrated isolation.

g. For Communications with a host, there is a wide variety of options to support operation in different locations. Standard daughter-boards can be mounted to support raw TCP/IP (most common), a web interface, RS232 or RS485 serial, USB, or Zigbee. 

Monitored data are averaged over an externally selectable interval. The interval is typically between 10s and 10min. The data for a period are sent as a single comma-delimited line. In addition, a copy of each line is stored in the SD card as backup.

\begin{figure}[h!]
\centering
\includegraphics[width=\linewidth]{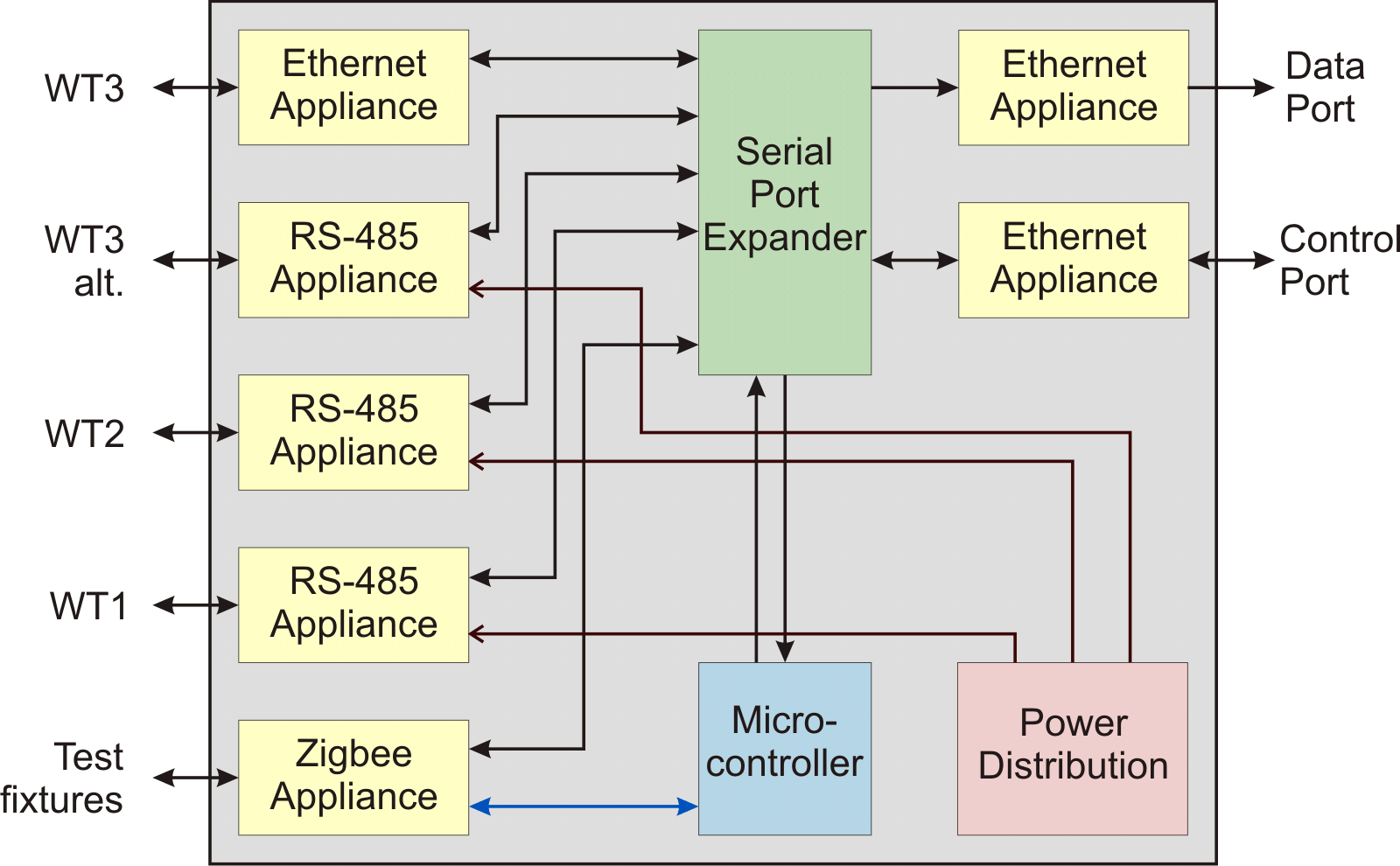}
\caption{Data Aggregator used during the 2011 deployment for collecting input from multiple SHMs then transmitting this data over Ethernet. \label{fig:pcu2011_block}}
\end{figure}

In most cases, the data are sent to a Data Aggregator which accepts streams from multiple SHMs and formats them as a single stream for further transmission. Figure (\ref{fig:pcu2011_block}) illustrates the Data Aggregator used during the 2011 deployment. 

\subsection{SSR Logic}
Control of the power flow uses the SSR bank. The logic is based on battery status and temperature following this outline:

a. If the battery temperature is below a selected Low-Temperature Threshold (LTT), input power is delivered to the heater, and the battery is disconnected. No power is delivered to the Load while the battery temperature is below the LTT. 

b. When the temperature exceeds the LTT and the battery voltage is below the Upper Voltage Threshold (UVT), input power is delivered to the battery.

c. If the temperature exceeds the LTT and the battery voltage exceeds the Lower Voltage Threshold (LVT), battery and input power are delivered to the Load.

d. If the battery voltage exceeds the UVT and the battery temperature is less than the Upper Temperature Threshold (UTT), the input power is routed to the heater.  

e. If the battery voltage exceeds the UVT and the battery temperature is greater than the UTT, the load will be connected to the battery, the heater will be off, and the input power will be directed to the Divert resistor.

f. All changes to the state of the battery connections include at least 1 volt of hysteresis.

\subsection{Remote Monitoring}
\begin{figure}[h!]
\centering
\includegraphics[width=\linewidth]{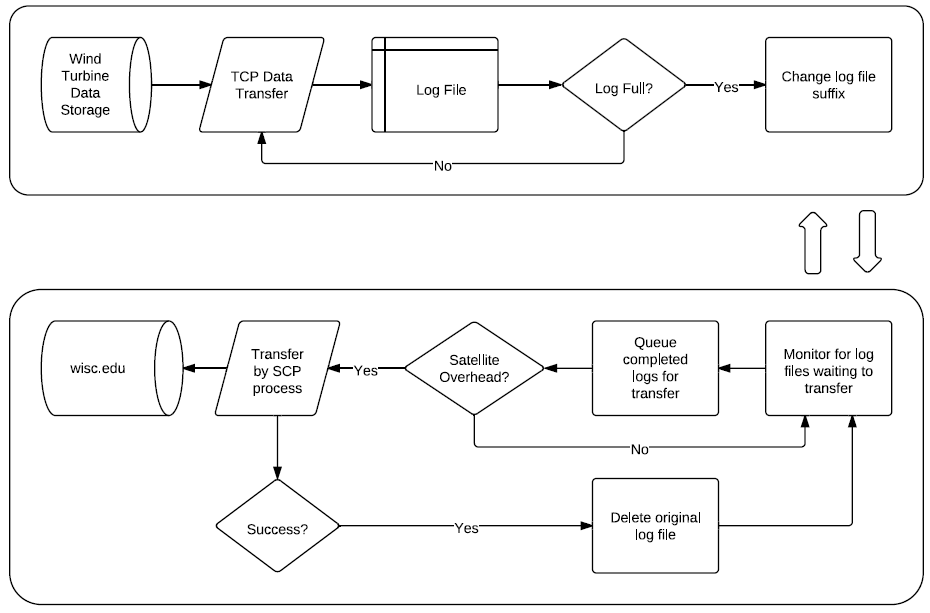}
\caption{Data storage and remote monitoring pipeline allowing remote analysis of turbine and PV panel performance as well as environmental sensors. \label{fig:flowpath}}
\end{figure}

A remote monitoring pipeline was designed to allow remote analysis of turbine and PV panel performance as well as environmental sensors. Sensor and performance data is continuously logged at regular intervals with logs stored locally on SD media and transmitted via satellite to an online database. Direct communication with the system health monitor allows remotely tuning control electronics and troubleshooting (Fig. {\ref{fig:flowpath}}).

\section{Modeling}
Computer modeling was used to simulate the operation of a station for different configurations of wind turbines and batteries. This was accomplished by creating models for wind speed, turbine power output, battery charging and discharging, and surface effects at different tower heights. These were ultimately combined to determine required levels of battery buffering in order to achieve a given target livetime.

\subsection{Wind Speed}
\begin{figure}[h!]
\centering
\includegraphics[width=0.8\linewidth]{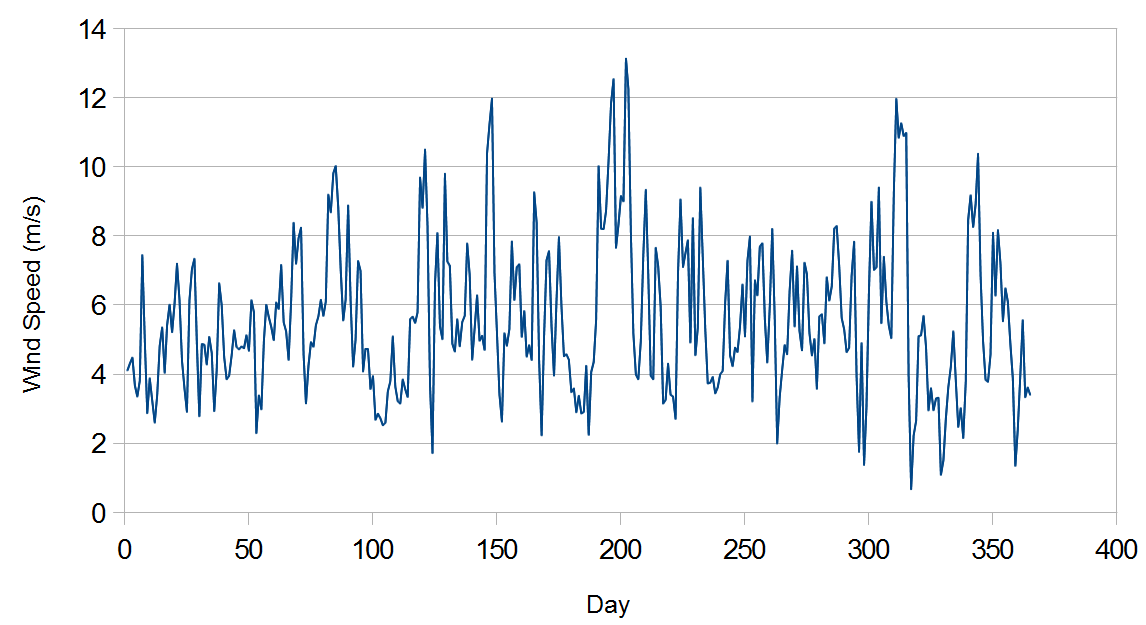}
\caption{Average wind speed at 10m height for South Pole Station from NOAA. \label{fig:avg_ws_sps}}
\end{figure}

Wind speed data from NOAA \cite{noaa} was used for predicting turbine power output. The NOAA metrology data sets contain wind speeds at 2m, 10m, and 30m elevation above surface, wind direction, barometric pressure, relative humidity at 2m, temperatures at 2m, 10m, and 30m, as well as precipitation, all at one minute intervals. Average wind speeds for 2010 are illustrated in Figure (\ref{fig:avg_ws_sps}).

\begin{figure}[h!]
\centering
\includegraphics[width=0.8\linewidth]{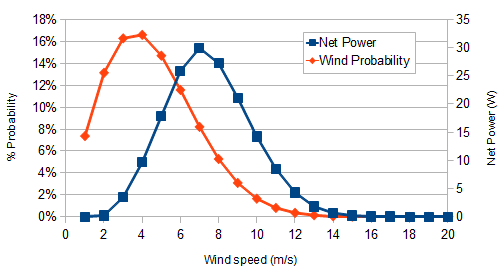}
\caption{Bergey XL1 Weibull distribution for predicting turbine power production. \label{fig:weibull}}
\end{figure}

The NOAA data was also used to construct a cumulative Weibull speed distribution for the purpose of predicting turbine power production. Figure (\ref{fig:weibull}) and Table (\ref{tb:weibull}) illustrate the results of this exercise. The sum of all the Net Power contributions is the average power output for the turbine over 24 hours. The primary result of our simulation is the prediction that, although the 120 Watt power requirement of the detector would be met on average, the wind turbine would only be above the cut-in threshold of 2--3 m/second approximately 80\% of the time. Therefore, battery buffering will be required.

\begin{table}[h!]
\centering
\begin{tabular}{| l | c |}
\hline
Hub average wind speed (m/s) & 4.55 \\
Air density factor & -2.8\% \\
Average output power (W) & 165 \\
Daily energy output (kWh) & 4.0 \\
Annual energy output (kWh) & 1,449 \\
Monthly energy output (kWh) & 121 \\
Percentage operating time & 78.8\% \\
\hline
\end{tabular}
\caption{Bergey XL1 Weibull Results. Although the 120 Watt power requirement of the detector would be met on average, the wind turbine would only be above the cut-in threshold of 2--3 m/second approximately 80\% of the time. Therefore, battery buffering will be required. \label{tb:weibull}}
\end{table}

\subsection{Power Output}
\begin{figure}[h!]
\centering
\includegraphics[width=0.8\linewidth]{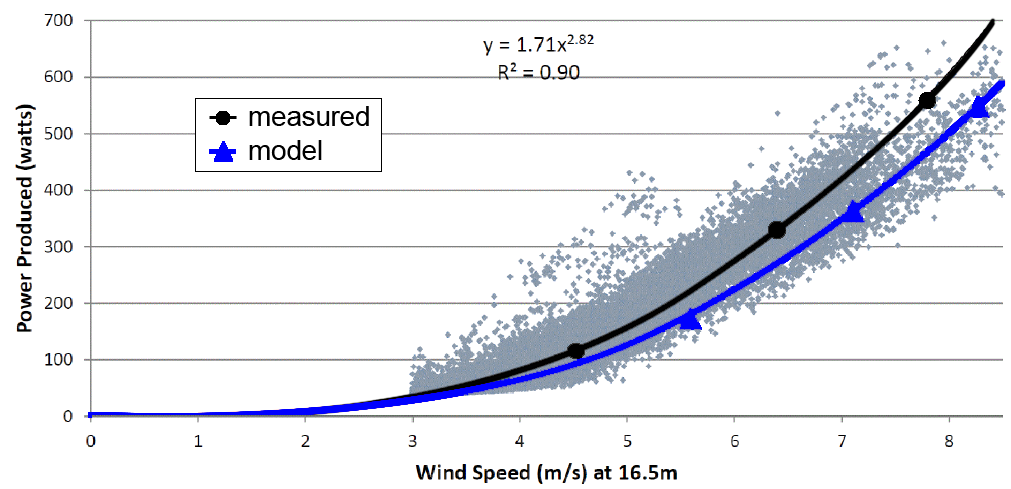}
\caption{Bergey XL-1 measured and modeled power output as a function of wind speed. The measured wind speed response roughly followed the expected cubic dependence of power on velocity. \label{fig:bergey_power}}
\end{figure}

Wind turbine manufacturer's power production curves were used along with a curve fitting tool \cite{zunzun} to find a continuous polynomial function to approximate the expected turbine power output. Figure (\ref{fig:bergey_power}) illustrates the measured power output for the Bergey XL-1 compared to the modeled power output. Software code was written for each of the candidate turbines for simulating power output as a function of wind speed. The measured wind speed response roughly followed the expected cubic dependence of power on velocity.

\subsection{Tower Height}
The NOAA wind speed data illustrated in Figure (\ref{fig:freq_dist}) indicates a reduction in wind speed at the surface boundary layer. Because wind power varies as the cube of wind speed, towers must be as high as possible to capture the highest wind energy. To predict wind speeds at the design tower heights, a tower height correction model was used to estimate wind speeds using wind direction and surface roughness data \cite{newcomb}. 

\begin{figure}[h!]
\centering
\includegraphics[width=0.7\linewidth]{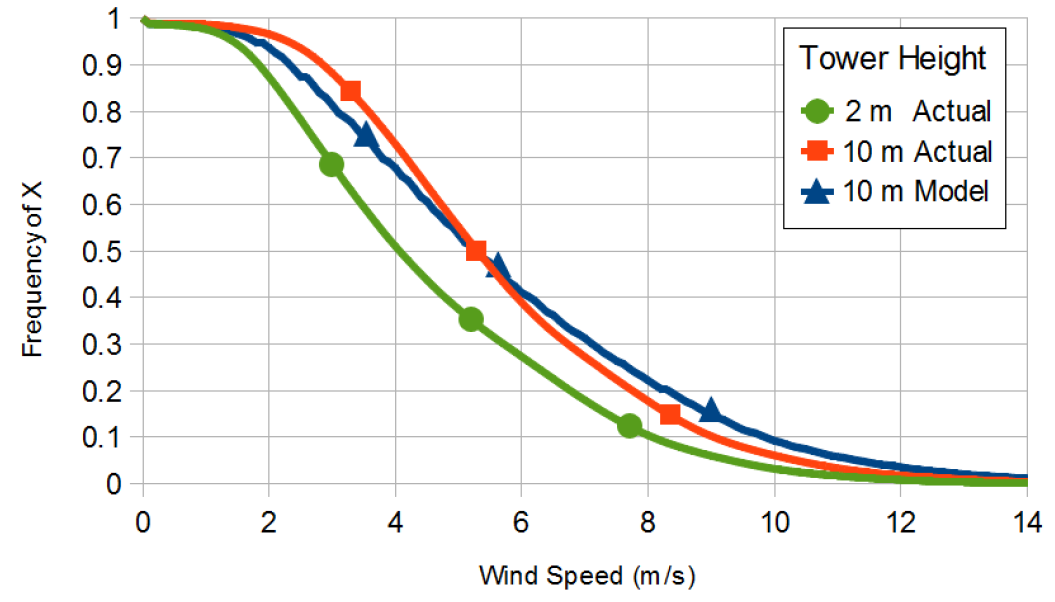}
\caption{Tower height correction model used to predict wind speeds at tower heights of 15 and 20m where NOAA data was not available. \label{fig:tower_model}}
\end{figure}

Figure (\ref{fig:tower_model}) illustrates the frequency distribution for NOAA wind data at 2m and 10m elevations (labeled $Actual$). The 10m $Model$ curve is the wind speed distribution at 10m as predicted by the model using the 2m elevation data. This model was used to predict wind speeds at tower heights of 15 and 20m where NOAA data was not available.

\subsection{Battery Buffering}

A model of battery charging and discharging characteristics was developed to determine the effect of battery buffering on station live-time. This included modeling battery capacity, voltage droop, and charge performance at different charging current levels for the Deka 8G31 12 Volt gel battery. 

\begin{figure}[h!]
\centering
\includegraphics[width=0.8\linewidth]{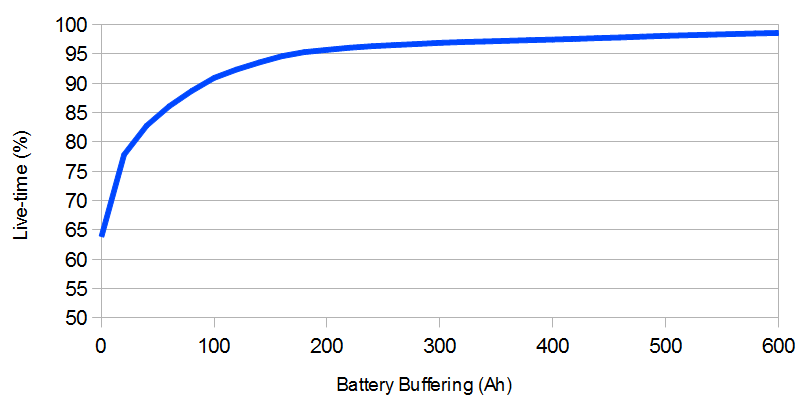}
\caption{2011 Percent live-time Bergey simulation. The simulation predicted that 98\% station live-time would be achieved for a 120W load and 600 Ah of buffering with ideal power output supplied from a single turbine and no PV panel. \label{fig:bergey_live_time}}
\end{figure}

Generator power output was simulated at the designated tower height using historical NOAA data with the tower height correction model. This simulation was performed using battery buffering levels of zero to 600 amp-hours, the Bergey turbine model, and wind speed data for the entire 2011 year. The simulation predicted that 98\% station live-time would be achieved for a 120W load and 600 Ah of buffering with ideal power output supplied from a single turbine and no PV panel (Fig. \ref{fig:bergey_live_time}).

\begin{figure}[h!]
\centering
\includegraphics[width=\linewidth]{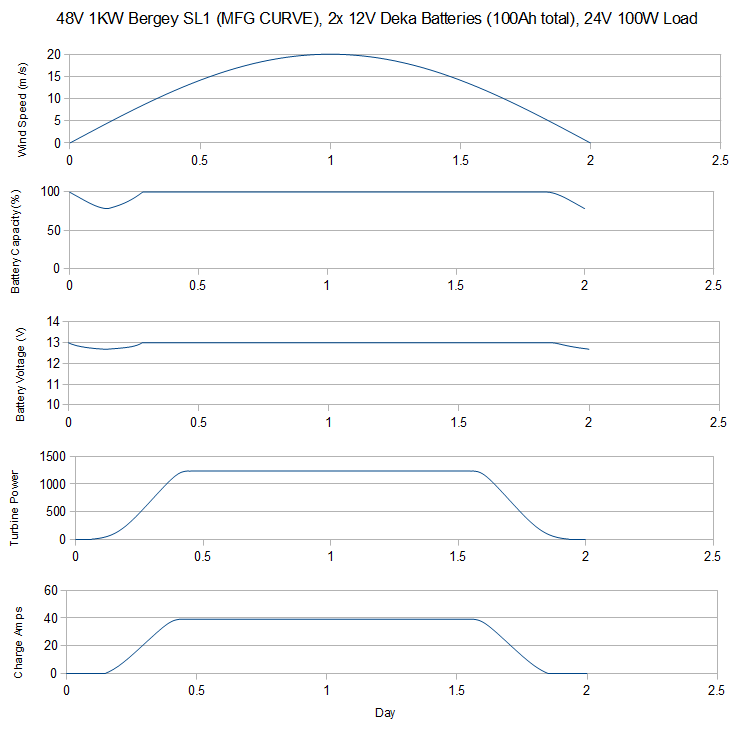}
\caption{Battery buffering model charging and discharging characteristics developed to determine the effect of battery buffering on station live-time. This included modeling battery capacity, voltage droop, and charge performance at different charging current levels for the Deka 8G31 12 Volt gel battery. \label{fig:batt_model_full}}
\end{figure}

Figure (\ref{fig:batt_model_full}) illustrates a simulated wind speed profile and the effect on battery capacity, battery voltage, turbine output power, and charging amps.

\section{Testing}
\subsection{Battery Performance}
\begin{figure}[h!]
\centering
\includegraphics[width=0.8\linewidth]{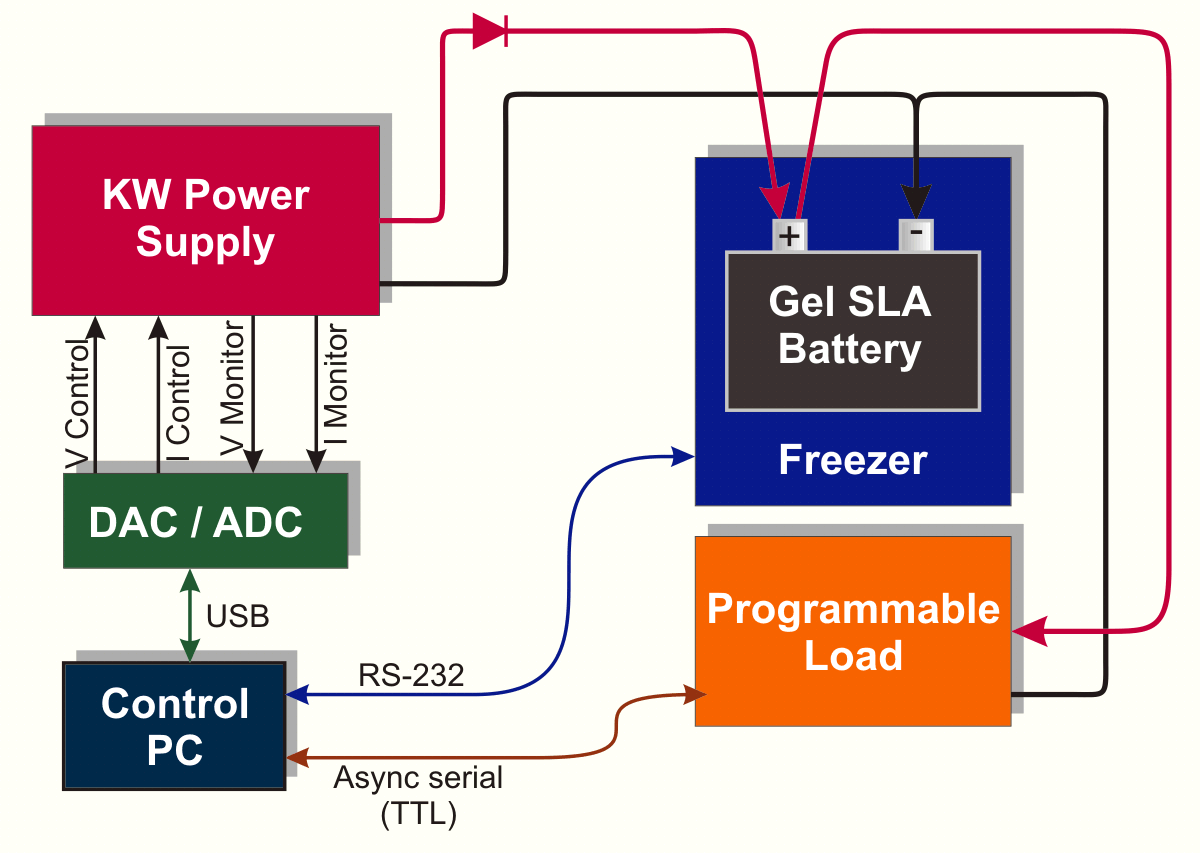}
\caption{Battery temperature test block diagram. A battery was charged and discharged at a series of temperatures to determine battery capacity as function of temperature. \label{fig:battery_test}}
\end{figure}

A battery was charged and discharged at a series of temperatures with measurements taken to determine battery capacity as a function of temperature. During this exercise, the thresholds for charge and discharge were adjusted for temperature. The block diagram for the experiment is shown in Figure \ref{fig:battery_test}.

\begin{figure}[h!]
\centering
\includegraphics[width=\linewidth]{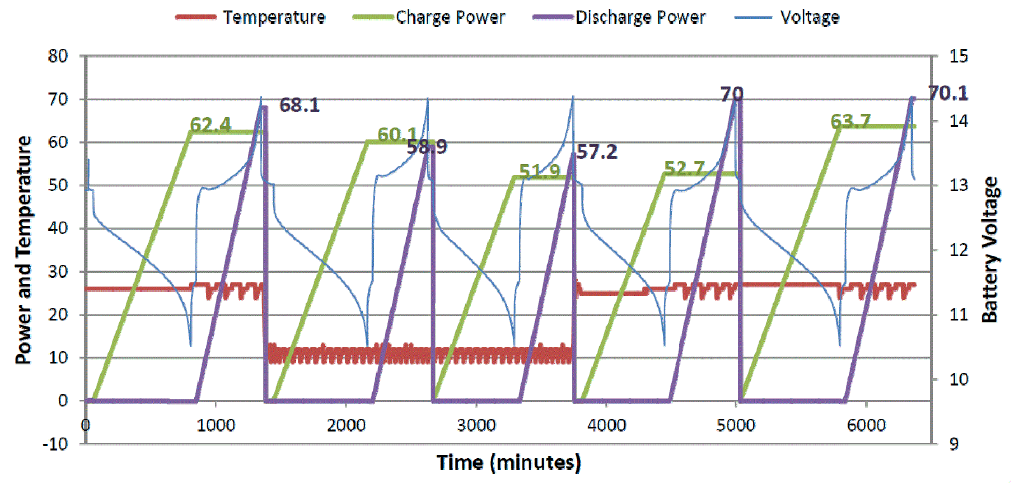}
\caption{Experimental results from battery testing. \label{fig:battery_profile}}
\end{figure}

Charging power was provided by a Sorensen 40-25 programmable power supply which was controlled and read out via a Measurement Computing USB 1208 DAQ board. The battery in the study was a 3 year old Deka 8G31 gel sealed lead acid battery. The battery was housed in a Tenney Jr environmental chamber. A BK 8512 Programmable Load was used to represent the in-field load and report the current and battery voltage. Typical experimental results are shown in Figure \ref{fig:battery_profile}.

\begin{figure}[h!]
\centering
\includegraphics[width=0.9\linewidth]{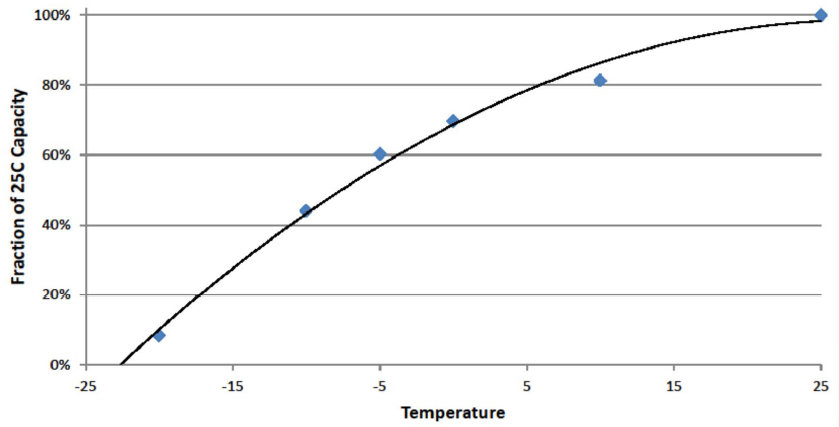}
\caption{Temperature profile of battery capacity. The batteries showed considerable degradation in performance at reduced ambient temperatures.  Consequently, heaters must be placed in the Instrument box and the SHM must route power to the heaters to ensure adequate battery performance. \label{fig:percent_capacity}}
\end{figure}

From these runs, the temperature profile of the battery capacity is plotted (Fig. \ref{fig:percent_capacity}) confirming that batteries show considerable degradation in performance at reduced ambient temperatures. Consequently, heaters must be placed in the Instrument box and the SHM must route power to the heaters to ensure adequate battery performance.

\subsection{Environmental Testing}
Environmental testing was performed using the Tenney Jr. environmental chamber with a Watlow controller. Candidate power cables were frozen to -80C and then bent to determine the allowable bending force and curvature limit before fracture. The cables must be flexible at low temperatures to allow installation without damage.

Wind turbines were disassembled to remove the generator bearings. The bearings were then tested at temperatures as low as -80C to ensure proper rotation. All bearings seized at temperatures below -80C using the stock grease. Multiple types of low temperature grease were tested as replacements. These included Royco 27, Molykote 33, and Mobil 33. All three types are currently in use at the deployed stations with no significant performance difference, although Mobil 33 will be retained for use in the future.

\section{Noise Measurements}
\begin{figure}[]
\centering
\includegraphics[width=0.8\linewidth]{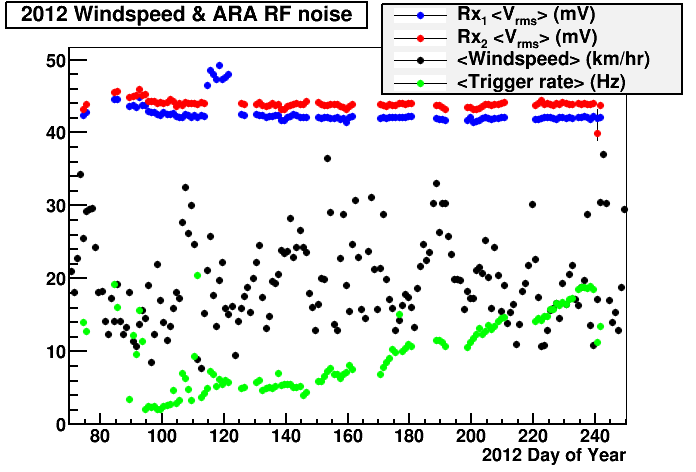}
\caption{Wind speed (black, in km/hr) compared to rms voltages (blue and red, respectively) for two surface antennas in proximity to wind turbine and associated monitoring and switching electronics.}
\label{fig:RFnoise}
\end{figure}

Since ARA is designed to detect the radio frequency signals caused by neutrino interactions in-ice, there was some concern regarding the possibility that the active electronics and switching electronics (relays) of the deployed wind turbines and control systems might radiate radio frequency backgrounds. We investigated possible correlations between wind speed and the average voltages recorded by two surface antennas, with frequency response from 25--300 MHz, in proximity (within 50 meters) of an Aero6gen turbine deployed in December, 2011. Over the course of several months data-taking in 2012, we do not observe any significant correlation between the rms voltage of those two surface antennas and the recorded South Polar wind velocities (Figure \ref{fig:RFnoise}, indicating that RF contamination is not large.

\message{WE NEED A SECTION ON RF NOISE GENERATED BY THE TURBINES WHICH I WILL SUPPLY.}

\section{Conclusions}
At the time of this writing, three successful deployments have been completed. The photovoltaic panel performed higher than expected. The Hummer turbine performed extremely well, however it also developed a vibrational instability that completely disabled the turbine within weeks of deployment. The Raum turbine was considered completely unsuitable for the environment, while the Bergey and Aero6gen turbines have been operating continuously without issue. Overall, power at a single station was proved extensible by installing multiple turbines and a PV panel.

The System Health Monitor was successfully used to collect measurements and make assessments of turbine and solar energy for use on the plateau. The communications pipeline, SHM, and environmental sensors have allowed remote monitoring and troubleshooting of system faults three years after initial deployment. Continued data collection and analysis has provided a window into system performance and directly contributed to continued development and preparation for future deployments.

Due to snow accumulation, having the instrumentation boxes buried in the ice  compounded maintenance and upgrade difficulty. Additionally, in the event that hardware problems could not be remotely troubleshooted, on site personnel would have to dig to get access to the SHM. This becomes logistically prohibitive as the number of stations increases and the snow accumulates each year.

\section{Extended Work}
The work on this project has been extended to several other projects including the Telescope Array RAdar project, TARA, and Antarctic Ross Iceshelf ANtenna Neutrino Array, ARIANNA, and a feasibility study for turbines deployed to Lake-62, Antarctica.

\subsubsection{Lake-62 Feasibility Study}

The wind turbine power output and battery buffering simulations developed during this project were used to determine the feasibility of deploying wind turbines at Lake-62, near Dome-A, Antarctica. Thirteen turbines were simulated with power outputs ranging from 4 to 15 kW. The battery charge and discharge model was used to predict live time percentage for a 4 kW load.

\begin{figure}[h!]
\centering
\includegraphics[width=\linewidth]{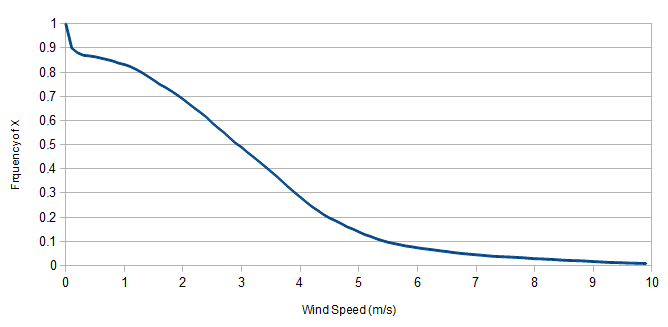}
\caption{Modeled wind frequency for Dome-A. Due to incomplete automatic wind station data for Lake-62, a model was developed to extrapolate wind conditions based on nearby station readings. \label{fig:lake62_wind}}
\end{figure}

Due to incomplete automatic wind station data for Lake-62, a model was developed to extrapolate wind conditions based on nearby station readings. This study revealed that the average wind speed at this location is less than the cut in speed for most wind turbines. Wind speeds were predicted to be above a cut in speed of 4 m/s only 30\% of the year (Fig. \ref{fig:lake62_wind}). Therefore, simulations were run for up to three turbines at the site such that the combined output would satisfy the 4 kW continuous load.

This study revealed that high kW wind turbines with significant battery buffering would be required to satisfy the continuous 4 kW load demand. Additionally, the 95\% live time drastically increased when using multiple high KW generators.

The following tables (\ref{tb:lake62_one_turb} - \ref{tb:lake62_three_turb}) list the turbine and battery configurations that would support 95\% live time.

\begin{table}[h!]
\centering
\caption{95\% Live-time for 1 turbine supplying 4kW load. \label{tb:lake62_one_turb}}
\begin{tabular}{| c | c | c | }
\hline
Turbine Model & Power Rating (kW) & Battery Buffering (Ah)\\
\hline
\hline
Gaia & 11 & 700 \\
Proven & 15 & 1175 \\
Aircon & 10 & 1450 \\
Aeolos & 10 & 1475 \\
\hline
\end{tabular}
\end{table}

\begin{table}[h!]
\centering
\caption{95\% Live-time for 2 turbines supplying 2kW load each. \label{tb:lake62_two_turb}}
\begin{tabular}{| c | c | c | }
\hline
Turbine Model & Power Rating (kW) & Battery Buffering (Ah)\\
\hline
\hline
Gaia & 11 & 275 \\
Aircon & 10 & 300 \\
Proven & 15 & 300 \\
Aeolos & 10 & 400 \\
Fortis Alize & 12 & 475 \\
Winforce & 10 & 550 \\
ARE 442 & 10 & 650 \\
Alize & 10 & 800 \\
Bergey XLS & 10 & 800 \\
Aeolos & 5 & 1025 \\
Evance & 5 & 1500 \\
\hline
\end{tabular}
\end{table}

\begin{table}[h!]
\centering
\caption{95\% Live-time for 3 turbines supplying 1.3kW load each. \label{tb:lake62_three_turb}}
\begin{tabular}{| c | c | c | }
\hline
Turbine Model & Power Rating (kW) & Battery Buffering (Ah)\\
\hline
\hline
Gaia & 11 & 100 \\
Aircon & 10 & 125 \\
Proven & 15 & 125 \\
Aeolos & 10 & 175 \\
Fortis Alize & 12 & 200 \\
Winforce & 10 & 275 \\
ARE 442 & 10 & 350 \\
Alize & 10 & 350 \\
Bergey XLS & 10 & 350 \\
Aeolos & 5 & 425 \\
Evance & 5 & 650 \\
\hline
\end{tabular}
\end{table}

\clearpage
\subsubsection{Telescope Array RAdar project, TARA}
\begin{figure}[h!]
\centering
\includegraphics[width=0.8\linewidth]{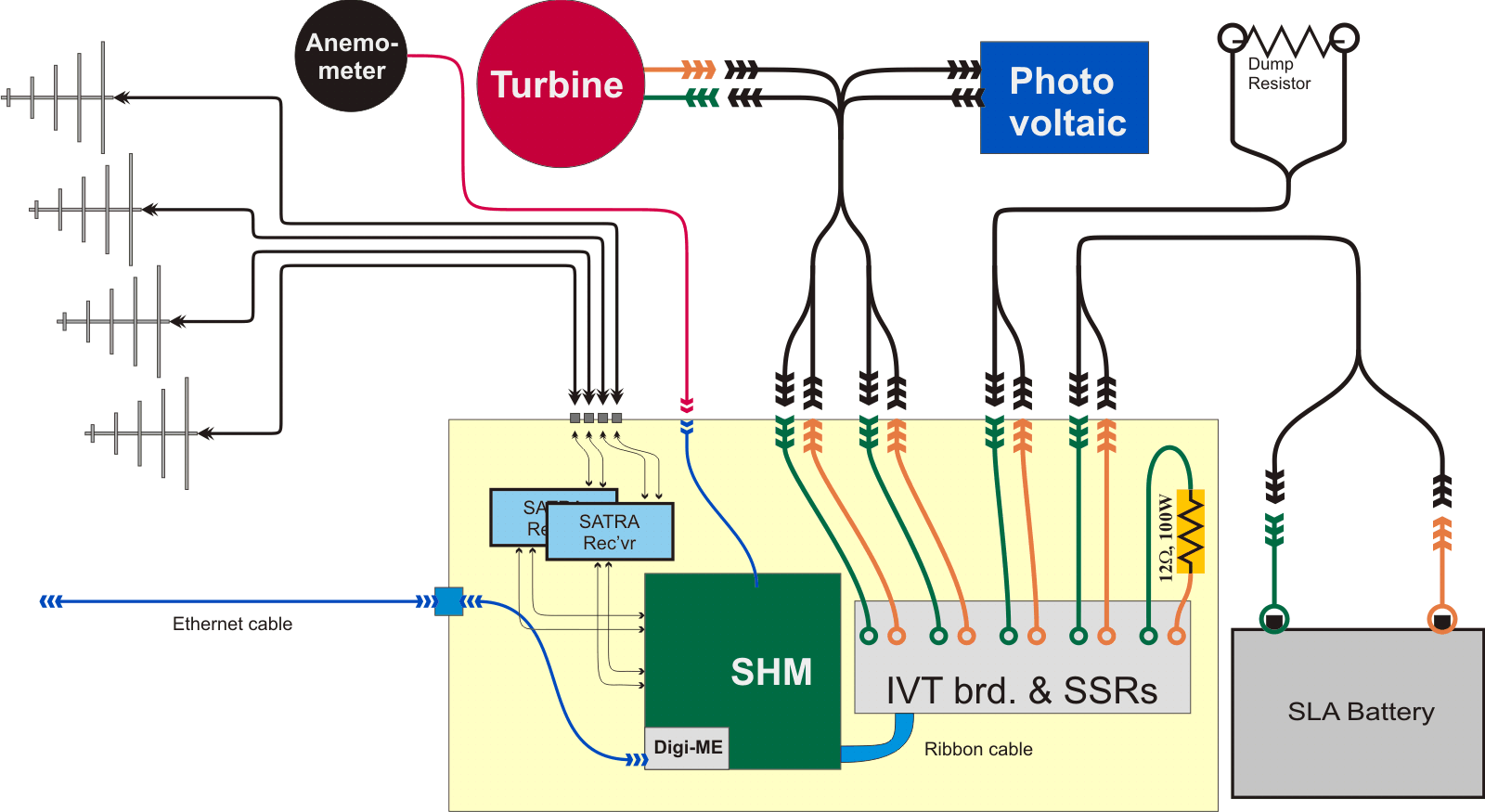}
\caption{A system health monitor was designed and built for this project by the Instrumentation Design Laboratory. Power is supplied by an Aero6gen turbine and a PV panel to remote TARA stations that consume a maximum power of 18 Watts. \label{fig:tara_block_diagram}}
\end{figure}

The TARA project utilizes remote stations with a transmitter and forward scattering radar to collect measurements on cosmic ray events. A system health monitor was designed and built for this project by the Instrumentation Design Laboratory. Power is supplied by an Aero6gen turbine and a PV panel to a maximum 18 Watt load (Fig. \ref{fig:tara_block_diagram}). 

\begin{figure}[h!]
\centering
\includegraphics[width=0.8\linewidth]{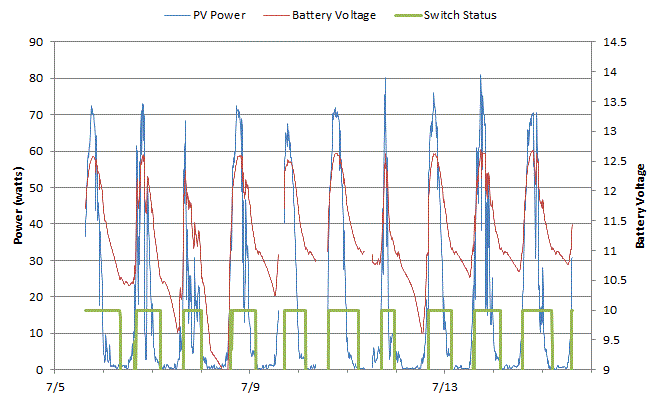}
\caption{TARA remote power charge and discharge data. The square wave is a relay signal that determines load switching and battery charging. The figure shows that a single PV panel was not able to maintain the batteries charged with a continuous 18 Watts discharge over a 24 hour cycle. \label{fig:tara_charge_discharge}}
\end{figure}

Figure (\ref{fig:tara_charge_discharge}) illustrates the data collected during the initial station deployment. At this point, power is supplied by the PV panel only. The square wave is a relay signal that determines load switching and battery charging. Since the station is designed to be powered primarily by wind, the figure shows that the PV panel cannot maintain the batteries charged with a continuous 18 Watts discharge over a 24 hour cycle.

\begin{figure}[h!]
\centering
\includegraphics[width=\linewidth]{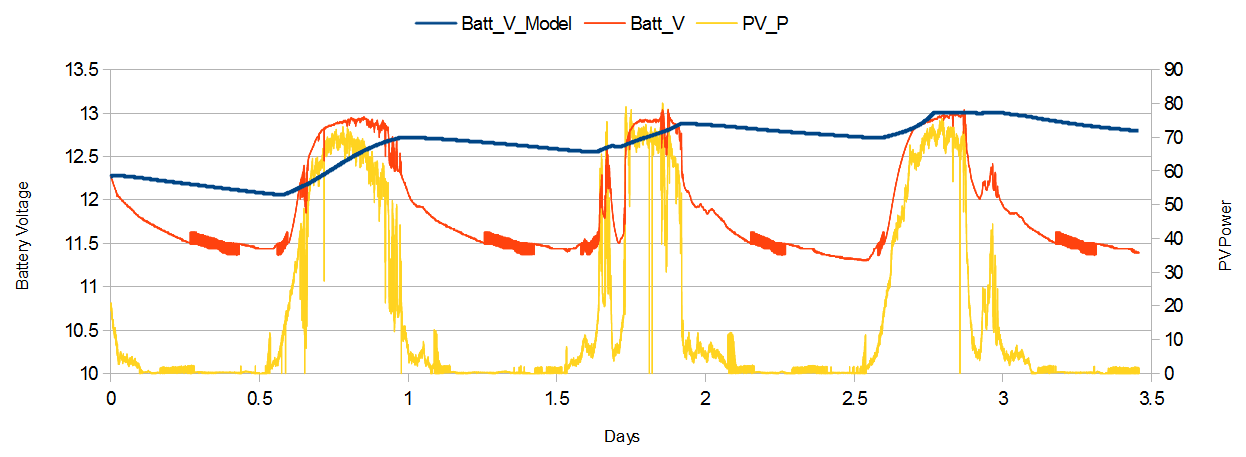}
\caption{TARA Power Simulation. This figure illustrates a comparison to three days of measured station data from August 2013 compared to modeled performance. The under-performance of the battery is believed to be due to battery failure as well as repeated full discharge cycling due to under-powering the system during the non-summer months. \label{fig:tara_sim}}
\end{figure}

The modeling discussed in Section (3) was adapted to compare theoretical performance of the TARA power configuration verses measured data. This model used measured PV panel power as input while simulating battery charging and discharging as well as load switching. After several months of deployment, the station batteries would no longer maintaining charge when solar power was not available. Figure \ref{fig:tara_sim} illustrates a comparison to three days of measured station data from August 2013 compared to modeled performance. The under-performance of the battery is believed to be due to battery failure as well as repeated full discharge cycling due to under-powering the system during the non-summer months. Future work will involve simulating different turbine, PV panel, and battery configurations for optimum station live-time.

\clearpage
\subsubsection{Antarctic Ross Iceshelf ANtenna Neutrino Array, ARIANNA}

\begin{figure}[h!]
\centering
\includegraphics[width=0.5\linewidth]{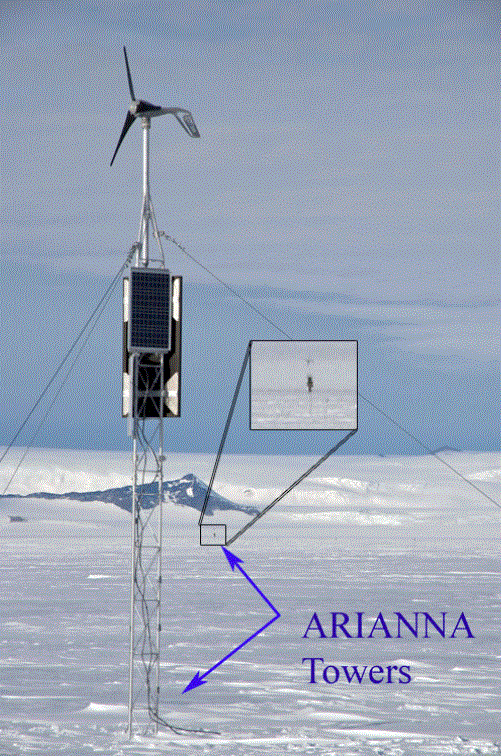}
\caption{ARIANNA power station. Figure courtesy of S. Barwick, UC, Irvine. \label{fig:arianna_turbine}}
\end{figure}

For the ARIANNA project, the Instrumentation Design Laboratory used the method explained in Section (4.2) to prepare wind turbines for deployment to the Ross Ice Shelf, Antarctica. Figure (\ref{fig:arianna_turbine}) shows two such turbines with PV panels in the field.

\section{Acknowledgments}
The authors would like to acknowledge the support of the National Science Foundation under grant NSF OPP-1002483 in addition to technical and logistical support provided by Ratheon Polar Services Corporation, Antarctic Support Contract, South Pole Station, and Ice-Cube personnel.

\clearpage

\bibliographystyle{model1-num-names}
\bibliography{mybib}{}

\end{document}